\newif\iffigs\figsfalse
  \figstrue
\iffigs \input epsf \else  \fi

\def\be{\begin{equation}}
\def\ee{\end{equation}}
\def\ba{\begin{eqnarray}}
\def\ea{\end{eqnarray}}
\def\smdr{\hbox{\hskip2pt\vrule height 4.1pt depth
-.3pt width .25pt \hskip-2pt$\times$}}
\def\NPB#1#2#3{{\sl Nucl. Phys.} \underbar{B#1} (#2) #3}
\def\PLB#1#2#3{{\sl Phys. Lett.} \underbar{#1B} (#2) #3}
\def\PRL#1#2#3{{\sl Phys. Rev. Lett.} \underbar{#1} (#2) #3}

\def\til{\widetilde}
\def\bar{\overline}
\def\bo{\hbox{1\kern -.23em {\rm l}}}
\def\bC{{\Bbb C}}

\def\bZ{{\Bbb Z}}
\def\bI{{\bf I}}
\def\bS{{\bf S}}
\def\bT{{\bf T}}
\def\CF{{\cal F}}
\def\CL{{\cal L}}

\def\CN{{\cal N}}
\def\CO{{\cal O}}

\def\tQ{{\til Q}}
\def\tq{{\til q}}
\def\ha{\mu}
\def\hb{\nu}
\def\t{{}^t\!}
\def\p{\partial}
\def\vev#1{{\langle #1\rangle}}
\documentstyle[aps,amsfonts,preprint,prl,floats]{revtex}

 \stop \fi

\relax
\begin{document}

\newlength{\pubnumber} \settowidth{\pubnumber}{IASSNS-HEP-94/999}
\preprint{\parbox{\pubnumber} {\begin{flushleft}
IASSNS-HEP-95/32\\
UK-HEP/95-06\\
hep-th/9505100\\ \end{flushleft}}}
\title{The Coulomb Phase of N=2 Supersymmetric QCD}
\author{Philip C. Argyres\footnote{argyres@guinness.ias.edu}
and M. Ronen Plesser\footnote{plesser@guinness.ias.edu}}
\address{School of Natural Sciences\\ Institute for Advanced Study\\
Princeton, NJ \  08540}
\author{Alfred D. Shapere\footnote{shapere@amoeba.pa.uky.edu}}
\address{Department of Physics and Astronomy\\ University of Kentucky\\
Lexington, KY \ 40506}
\maketitle
\begin{abstract}
We present an explicit non-perturbative solution of $\CN{=}2$
supersymmetric $SU(N)$ gauge theory with $N_f \le 2N$ flavors
generalizing results of Seiberg and Witten for $N=2$.
\end{abstract}

In this letter we construct an effective Lagrangian describing the
low-energy dynamics of $\CN{=}2$ supersymmetric $SU(N)$ QCD with
$N_f=2N$ flavors in its Coulomb phase.  Being exact,
this solution makes possible the study of strong--coupling phenomena in
the theory.  The solution exhibits exact scale invariance and
strong-weak coupling duality.
All asymptotically free $\CN{=}2$ $SU(N)$ theories with
$N_f < 2N$ are obtained as appropriate limits of this solution in
parameter space.  Similar ideas have been successfully applied to
the case $N=2$;  see \cite{SWii} and references therein.

$\CN{=}2$ QCD is described in terms of $\CN{=}1$ superfields by a chiral
field strength multiplet $W^a_b$ and a chiral multiplet $\Phi^a_b$ both
in the adjoint of the gauge group, and chiral multiplets $Q^i_a$ in the
$\bf N$, and $\tQ^a_i$ in the ${\bf \bar N}$ representations of the
gauge group.  Flavor and color indices are respectively
$i,j,k=1,\ldots,N_f$  and $a,b,c=1,\ldots,N$.

The Lagrangian contains $\CN{=}1$ gauge-invariant kinetic terms for the
fields with gauge coupling constant $\tau =
(\theta/\pi) + i (8\pi/g^2)$, and the superpotential
		\be
	{\cal W}= \sqrt2
\tQ^a_i \Phi_a^b Q_b^i + \sqrt2 M^i_j \tQ^a_i Q_a^j.
	\ee
The quark mass
matrix satisfies $[M,M^\dagger]=0$ implying that it can be diagonalized
by a flavor rotation to $M = {\rm diag}(m_1,\ldots,m_{N_f})$.
Classically and with vanishing masses, the global symmetries are a
$U(N_f)$ flavor symmetry and a $U(1)_R \times SU(2)_R$ chiral
R-symmetry.  The trace of the mass matrix $M$ is a flavor singlet,
while the rest transforms in an adjoint flavor representation.  We
denote the flavor-singlet and flavor-adjoint masses by, respectively,
\be\mu\equiv (1/N_f) \sum_{j=1}^{N_f} m_j, \qquad \mu_j \equiv m_j -\mu.\ee

The theory has a rich vacuum structure.  In this letter we study its
Coulomb phase, where the vevs of the lowest components of the chiral
superfields satisfy $q^i_a=\tq^a_i=0$ and $[\phi,\phi^\dagger]=0$.
This implies that $\phi$ can be diagonalized by a color rotation to a
complex traceless matrix $\vev{\phi} = {\rm
diag}(\phi_1,\ldots,\phi_{N}).$
As gauge--invariant coordinates on the Coulomb phase moduli space, we
take the elementary symmetric polynomials $s_\ell$ defined by
\be{\rm det}(x-\vev\phi)=\prod_a(x-\phi_a) = \sum_{\ell=0}^{N} s_\ell
x^{N-\ell}.\ee
Note that $s_0=1$ and $s_1=0$ by the tracelessness of $\phi$.

$\vev\phi$ generically breaks the gauge symmetry $SU(N)$ to
$U(1)^{N-1}$ and gives all the quarks masses, so the low energy
effective theory is an $\CN{=}2$ supersymmetric $U(1)^{N-1}$ Abelian
gauge theory.  If some of the $\phi_a$'s are equal, the unbroken gauge
group will include non-Abelian factors.  Also, when $\phi_a + m_i=0$ a
quark is massless.

If we assume that $\CN{=}2$ supersymmetry is not dynamically broken,
then the Coulomb vacua are not lifted by quantum effects.  At a
generic point, the low energy effective Lagrangian can be written in
terms of the $\CN{=}2$\ \ $U(1)$ gauge multiplets $(A_\ha,W_\ha)$,
where $\ha,\hb=1,\ldots,N{-}1$ and label quantities
associated to each of the $U(1)$ factors.  We denote the scalar
component of $A_\ha$ by $a_\ha$, which we will also take to
stand for its vev.

The $\CN{=}2$ effective Lagrangian is determined
by an analytic prepotential $\CF(A_\ha)$ and takes
the form
		\be
	\CL_{\rm eff} = {\rm Im}{1\over4\pi} \left[
	\int\!\! d^4\theta\, A_D^\ha\,\bar{A_\ha} +
	{1\over2}\int\!\!d^2\theta\,\tau^{\ha\hb}\,W_\ha W_\hb \right] ,
		\label{effL}\ee
where the dual chiral fields and the effective couplings are given by
\be A_D^\ha \equiv {\partial\CF/\partial A_\ha}, \qquad \tau^{\ha\hb}
\equiv {\partial^2\CF/\partial A_\ha\partial A_\hb}.\ee
Typically,
this effective action is good for energies up to the mass of
the lightest massive
particle.  There are special submanifolds of moduli space where extra
states become massless.  As we approach these submanifolds the range of
validity of (\ref{effL}) shrinks to zero; on these singular
submanifolds the effective Lagrangian must be replaced with one which
includes the new massless degrees of freedom.

The $U(1)^{N-1}$ theory has a lattice of allowed electric and
magnetic charges, $q^\ha$ and $h_\ha$.  Generically the bare masses
break the flavor symmetry $U(N_f) \rightarrow U(1)^{N_f}$, so states
will have associated quark number charges $n^j\in\bZ$.  A BPS saturated
$\CN{=}2$ multiplet with quantum numbers $q^\ha$, $h_\ha$, and $n^j$ will
have a mass given by the central charge formula \cite{SWii}
\be
M = |a_\ha q^\ha +
a^\ha_D h_\ha + m_j n^j|.\ee

The physics described by the $U(1)^{N-1}$ effective theory is
invariant under an $Sp(2N{-}2;\bZ)\smdr \bZ^{N_f}$ group of duality
transformations, which acts on the scalar fields and their duals, as
well as the electric, magnetic, and quark number charges, in such a
way as to leave the central charges invariant.
Encircling a singular submanifold in moduli space
may produce a non-trivial duality
transformation.

More explicitly, consider $\bS\in Sp(2N{-}2,\bZ)$ and $\bT$ a
$(2N{-}2)\times N_f$ integer matrix.  Then a duality transformation
$(\bS,\bT)$ acts on the fields and charges as ${\bf a} \rightarrow
\bS\cdot{\bf a}+ \bT\cdot{\bf m}$,\ \ ${\bf h} \rightarrow
\t\bS^{-1}\cdot{\bf h}$, and ${\bf n} \rightarrow -\bT\cdot{\bf h} +
{\bf n}$.  Here we have defined the column vectors $\t{\bf a} \equiv
(a_D^\ha, a_\hb)$, $\t{\bf m} \equiv (m_j)$, $\t{\bf h} \equiv
(h_\ha,q^\hb)$, and $\t{\bf n} \equiv (n^j)$.

In a vacuum with massless charged particles, the $U(1)$ gauge fields
that couple to them will flow to zero coupling in the infrared and will
be well-described by perturbation theory.  Thus a 1-loop calculation
suffices to determine the monodromy around a submanifold of such vacua
where one dyon with charges $({\bf h},{\bf n})$ is massless.  The
monodromy around such a singularity is
                \ba
       \bS &&= \bo+{\bf h}\otimes\t(\bI\cdot{\bf h})
	= \pmatrix{\delta^\ha_\hb + q^\ha h_\hb & q^\ha q^\hb\cr
	- h_\ha h_\hb & \delta^\hb_\ha - h_\ha q^\hb},\nonumber\\
	{}&&\nonumber\\
	\bT &&= {\bf n}\otimes\t(\bI\cdot{\bf h})
	= \pmatrix{ n^j q^\ha \cr -n^j h_\hb},
	\label{dyonmon}\ea
where $\bI={0\ \ 1\choose -1\ 0}$ is the symplectic metric.

\paragraph*{Solving the theory.}

Our aim is to determine the analytic prepotential $\CF$ of the
low-energy Abelian theory everywhere on moduli space.  We will assume,
as in \cite{SWii,SWi,AF,KLTY}, that the effective coupling
$\tau^{\ha\hb}(s_\ell)$ is the period matrix of a genus $N{-}1$ Riemann
surface $\Sigma(s_\ell)$ varying holomorphically over moduli space.  We
solve the theory by constructing this family of surfaces.

Concretely, the construction requires a family of curves as above and a
meromorphic form $\lambda (s_\ell)$ on $\Sigma (s_\ell)$ satisfying
		\be
	{\p\lambda\over\p s_\ell} = \omega_\ell + d f_\ell,
	\qquad \ell = 2, \ldots, N,
		\label{difeq}\ee
with $\omega_\ell$ a basis of $N{-}1$ holomorphic one-forms on
$\Sigma$, and $f_\ell$ arbitrary functions, and with residues at its
poles which are integral linear combinations of $m_j$, the bare quark
masses. Choosing a basis of $2N{-}2$ one-cycles
$(\alpha^\ha,\beta_\hb)$ on $\Sigma$ with the standard intersection
form $\vev{\alpha^\ha,\beta_\hb} = \delta^\ha_\hb$,
$\vev{\alpha^\ha,\alpha^\hb} = \vev{\beta_\ha,\beta_\hb}=0$, the vevs
of the scalar fields and their duals are defined by
		\be
	a_D^\ha = \oint_{\alpha^\ha} \lambda, \qquad
	a_\hb = \oint_{\beta_\hb} \lambda.
		\ee
This defines the vevs up to duality transformations
$(\bS,\bT)$, an ambiguity corresponding to our freedom to choose a
symplectic basis ($\bS$) and to shift the winding numbers of the cycles
around each of the poles ($\bT$).  The condition on the residues of
$\lambda$ guarantees that the correct action of $\bT$ on the vevs is
realized.  Physically, the $\bT$ ambiguity in the periods corresponds
to the freedom to shift the global quark-number current by a multiple
of a $U(1)$ gauge current \cite{SWii}.

We will further assume, as in \cite{AF,KLTY}, that $\Sigma$ is a
hyperelliptic Riemann surface with polynomial dependence on the
coordinates $s_\ell$ and the masses $m_j$. A curve $y^2= \wp(x)$ where
$\wp(x)$ is a polynomial in $x$ of degree $2N$, describes a
hyperelliptic Riemann surface of genus $N{-}1$ as a double-sheeted
cover of the $x$-plane branched over $2N$ points.

We now proceed to determine the curve and one-form for the $N_f=2N$
theory using this {\it Ansatz}.  First consider the $SU(2N)$ theory
with $2N$ flavors.  This theory has a dynamically-generated scale which
we denote $\Lambda$.  Let us denote the coordinates on the moduli space
by the usual symbols with tildes.  Recall that in this theory the
$U(1)_R$ symmetry is broken by instantons to a $\bZ_{2N}$ under which
$\til\phi_a$, $\til m_j$, and $\Lambda$ all have charge 1.  As in
\cite{AF,KLTY}, we choose charges 1 and $2N$ for $x$ and $y$ so that
this is a symmetry of the curve. With these assignments, $\wp$ must be
a homogeneous polynomial in $x$, $\til\phi_a$, $\til m_j$, and
$\Lambda$ of degree $4N$.  Since $\Lambda$ must enter with its
instanton weight $\Lambda^{2N}$, we can can write $\wp= \til P -
\Lambda^{2N} \til Q + \Lambda^{4N} \til R$, where $\til P$ is of degree
$4N$ in $x$, $\til\phi_a$, and $\til m_j$; $\til Q$ is of degree $2N$;
and $\til R$ is a constant.

The weak coupling limit is $\Lambda\rightarrow 0$.  In this limit the
branch points of the curve are at the zeros of $\til P$. When two (or
more) of these coincide one or more cycles of the Riemann surface
degenerate. This, in turn, corresponds by our construction of the vevs
as period integrals and by the central charge formula, to some charged
state becoming massless.  Such massless states appear at weak coupling
wherever two of the eigenvalues of $\til\phi$ coincide; $\til P$ has
double zeros in $x$ for these values of $\til\phi$.  Since the
$\til\phi_a$'s can only enter symmetrically (for gauge invariance), we
must have $\til P = \hat P\cdot\prod_{a=1}^{2N} (x-\til\phi_a-\til\mu)$
where $\til\mu$ is some fixed linear combination of the $\til m_j$ and
$\hat P$ is a homogeneous polynomial of degree $2N$.  Furthermore, the
whole curve must be singular as $\Lambda\rightarrow 0$, in order to
produce the degeneration corresponding to the $2N$-flavor $SU(2N)$ beta
function.  Thus, at least one of the zeros of $\hat P$ must in fact be
at $\til\phi_a+\til\mu$ for some $a$.  Then, by symmetry in the
$\til\phi_a$'s and using our freedom to rescale $y$ and shift and
rescale $x$, we find $\til P = \prod_{a=1}^{2N}(x-\til\phi_a)^2$.

Now let us break $SU(2N)\rightarrow SU(N)\times SU(N)^\prime \times
U(1)$ at a large mass scale and tune the bare masses so that there are
$2N$ light hypermultiplets transforming as $({\bf N},{\bf 1},0)$.  This
is achieved by setting $\til\phi_a = M + \phi_a$ for $a=1,\ldots,N$,
and $\til\phi_a = -M + \phi_a^\prime$ for $a=N+1,\ldots,2N$, with
$\sum_a \phi_a = \sum_a \phi_a^\prime = 0$.  We should also set $\til
m_j = -M + m_j$ to get quarks with masses $m_j$ in the first $SU(N)$
factor.  In the limit $M\gg (\phi_a, \phi_a^\prime, m_j)$ the three
factors decouple. To obtain the $SU(N)$ factor with $2N$ flavors at
finite coupling $\tau$, we should send $\Lambda \rightarrow \infty$
such that $(\Lambda/M)^{2N} \propto q \equiv e^{i\pi\tau}$.  Here we
have used the one-loop renormalization group matching.

For $\CN{=}2$ supersymmetric theories there are no higher-loop
corrections; however, non-perturbative (instanton) corrections may
modify the matching by higher powers of $q$.  Similarly, the classical
mass matching may be modified by quantum corrections.  Thus the
matching conditions in the limit $M\rightarrow\infty$ are
$(\Lambda/M)^{2N}=f(q)\propto q+\CO(q^2)$ and $m_j = -M + g(q) \mu +
h(q) \mu_j$, with $g(q), h(q) = 1 + \CO(q)$.
The two undetermined functions $g$ and $h$ arise from the two possible
renormalizations of the mass term: one for the flavor
singlet masses and one for the flavor adjoint masses.  Their leading
behavior in $q$ comes from comparing to the weak-coupling limit, $q
\rightarrow 0$.

Now take the limit $M\rightarrow\infty$ in the $SU(2N)$ curve with
$2N$ flavors.  After shifting $x \rightarrow x+M$, the polynomial
$\til P(x)$ factorizes into a piece with $2N$ zeros near $x=0$ (relative
to the scale $M$) and another piece with zeros all near $x=2M$.  The
whole curve $y^2=\til P+\ldots$ should factorize in this way to
correspond to the decoupling low-energy sectors.  Up to a
conformal transformation, this process describes the degeneration of
the original Riemann surface into two Riemann surfaces, corresponding
to the two decoupled $SU(N)$ factors, connected by two long
necks. Taking $x \ll M$ and rescaling $y \rightarrow (2M)^N y$, the
curve becomes approximately
		\be
	y^2 = P^2 - f\til Q(x+M, \til m_j, \til\phi_a)
	+ f^2 M^{2N} \til R ,
		\label{faccur}\ee
where $P\equiv \prod_{a=1}^N(x-\phi_a)$.

The factorization discussed above means this polynomial is independent
of the
$\phi'_a$.  Being a symmetric function of the $\til\phi_a$, $\til Q$
must in fact have no $\til\phi$ dependence. Factorization also requires that
the branch points of (\ref{faccur}) be at $|x|\ll M$, so the
coefficients of positive powers of $M$ must vanish identically in $x$
and $q$. Thus $\til R = 0$, while in terms of $g(q)\mu = \til\mu + M$
and $h(q)\mu_j = \til\mu_j$, $\til Q$ must take the form $\til
Q(x+g\mu,h\mu_j)$.

To further constrain the curve we construct the meromorphic one-form
$\lambda$.  A basis of holomorphic one-forms on our hyperelliptic curve
are $\omega_\ell = x^{N-\ell}dx/y$ for $\ell=2,\ldots,N$.  Noting that
$P=\sum_{\ell=0}^N s_\ell x^{N-\ell}$, it is straightforward to
integrate the differential equation (\ref{difeq}) to find the solution
$\lambda = a \log [(P-y)/(P+y)]dx$ where $a$ is independent of $x$ and
the $s_\ell$'s.  But this $\lambda$ has logarithmic singularities at
$x=\epsilon_j$, the zeros of $\til Q$.  These logs can be converted
into poles by adding the total derivative $d[a(x+b)\log ((P+y)/(P-y))]$
to $\lambda$. This
does not affect the differential equation (\ref{difeq}).  The resulting
form has poles $\pm a(\epsilon_j+b)dx/(x-\epsilon_j)$ at the two
preimages of $\epsilon_j$.

The requirement that the residues of $\lambda$ be linear
in the quark masses implies that the $\epsilon_j$, the zeros of $\til
Q$, are linear in the masses.  The most general flavor-symmetric
$\til Q$ with this
property is $\til Q=\prod_j (x+g \mu + h \mu_j)$. In fact one
renormalization of the masses, say $h(q)$, can be absorbed into the
definition of the coupling.  Henceforth we will set $h=1$.  Finally,
demanding that the residues of $\lambda$ be the bare
quark masses $\pm m_j$ implies $a=1/2\pi i$ and $b=(g-1)\mu$.

The resulting one-form has an additional pole at $x=\infty$ which is
easily calculated to be $\pm{2N}\mu{g}(1-f)^{-1/2}(dx/x)$.  As we
will show later, the residues of $\lambda$ on a given sheet of the
hyperelliptic curve ({\it i.e.}, choosing one branch of the square root
in $y$) must sum to zero in order to reproduce a scale-invariant
theory.  Since $2N \mu = \sum_j m_j$ we must have $f=1-g^2$ for this
vanishing condition to hold.

The curve is thus
		\ba
	y^2 &&= P^2 - f\, Q,\nonumber\\
	P &&\equiv \prod_{a=1}^{N} (x -\phi_a) ,\nonumber\\
	Q &&\equiv \prod_{j=1}^{2N} ( x + g\, \mu + \mu_j) ,\nonumber\\
		f &&= 1- g^2,
	\label{curvformiii}\ea
and the one-form is
		\be
	\lambda = {x+(g-1)\mu\over 2\pi i}\,
	d\left[ \ln\left(P-y \over P+y \right) \right] .
	\label{oneform}\ee

To complete the argument we need to determine the function $g(q)$.  In
principle $g$ could depend on $N$ as well as $q$.  We first determine
its $N$-dependence by induction in $N$, then determine the
$q$-dependence by matching onto the $N=2$ case.

The induction proceeds by considering the breaking of $SU(N)$ with
$N_f=2N$ down to $SU(N{-}1)\times U(1)$ with $2N{-}2$ light
hypermultiplets transforming as $({\bf N{-}1},0)$. Set
		\ba
	\phi_a &&= \left\{ \matrix{
	M+\phi '_a\hfill &\qquad\qquad\ \  a=1,\ldots,N{-}1,\hfill\cr
	(1-N)M\hfill &\qquad\qquad\ \  a=N,\hfill\cr} \right. \nonumber\\
	{}&&\nonumber\\
	m_j &&= \left\{ \matrix{
	-t(q)M+ m'_j \hfill&\qquad j=1,\ldots,2N{-}2,\hfill\cr
	-u(q)M + v(q)\mu ' \hfill&\qquad j=2N{-}1,2N.\hfill\cr} \right.
	\label{induc}\ea
The limit $M\rightarrow\infty$ keeping $\phi '_a$ and $m'_j$ fixed
achieves the desired breaking at weak coupling if $t(0)=1$ and $u(0)\ne
1$.  The functions $t$, $u$, and $v$ will
be determined by demanding that our curve (\ref{curvformiii})
reduces to a curve of the same form with $N\to N{-}1$.  Plugging
(\ref{induc})
into (\ref{curvformiii}) and taking the $M\rightarrow\infty$ limit, one
finds after some algebra that $f=1-g^2$ for any $N$ implies $g_{N-1} =
g_N$, and $f_{N-1} = f_N$.  The one-loop  renormalization group
matching condition that $f_N(q) \propto q$ independent of $N$ implies
that the bare couplings satisfy $\tau_N = \tau_{N-1}$ at weak
coupling.

We now determine the unknown function $g(q)$ by matching to the known
solution \cite{SWii} of the $SU(2)$ theory with 4 massless
flavors $\til y^2 =
\prod_{i=1}^3 ( \til x - e_i s_2 )$, where
	\be
3e_1 \equiv \theta_2^4 +
\theta_3^4,\qquad 3e_2 \equiv -\theta_3^4 - \theta_1^4, \qquad
3e_3 \equiv
\theta_1^4 - \theta_2^4.\ee
The theta functions are defined as in
\cite{SWii}.

This curve is equivalent to (\ref{curvformiii}) with
$\mu=\mu_j=0$ if there is an $SL(2,\bC)$
transformation relating the $x$ and $\til x$ coordinates which
maps the branch points of one curve into those of the
other.  This condition determines $f$ in terms of the $e_i$
up to cyclic permutations.  Since $g=\pm\sqrt{1-f}$, there are in all
six possible solutions, of which only two have the correct
weak-coupling asymptotic form $g=1+\CO(q)$:
		\be
	g = {\theta_2^4+\theta_1^4\over \theta_2^4-\theta_1^4},
	\qquad {\rm or} \qquad
	g = {\theta_3^4-\theta_1^4\over \theta_3^4+\theta_1^4}.
		\label{ganswr}
		\ee
These are physically equivalent: they are related by $\tau \rightarrow
\tau+1$, or equivalently, by a conventional choice of the origin of the
$\theta$ angle.  This completes the determination of the $N_f=2N$
curve.  The curves for $N_f<2N$ are obtained by taking appropriate
limits as $q\to 0$ and $m_j\to\infty$.  In particular, this reproduces
the results of \cite{AF,KLTY}.

\paragraph*{Properties of the solution.}
The modular functions (\ref{ganswr}), and therefore the
$N_f=2N$ curve, are invariant under $T^2 : \tau \rightarrow \tau+2$.
This is expected, since it corresponds to a shift in the theta angle
$\theta \rightarrow \theta + 2\pi$, which is a symmetry of the
theory.
In addition there is a strong--weak coupling duality.
Taking, for example, the first solution in (\ref{ganswr}), we
see that
$g \rightarrow -g$ under $S:\tau\rightarrow-1/\tau$.  The curve
(\ref{curvformiii}) is left invariant if, at the same time, we take
$\mu \rightarrow -\mu$ and $\mu_j \rightarrow +\mu_j$, verifying a
conjecture of ref.\ \cite{LS}.

As a check on the validity of our solution, we now show that it
reproduces the positions and monodromies of two classes of
singularities at weak coupling.

The first class, which we refer to as gauge singularities, occurs
whenever $\phi_a=\phi_b$.  These singularities correspond classically
to the restoration of a non-Abelian gauge symmetry.  Because the beta
function vanishes, the semiclassical monodromies around the gauge
singularities are actually the classical monodromies given by elements
of the Weyl group of $SU(N)$, which act by permuting the $\phi_a$'s.
The breaking (\ref{induc}) implies that all the $SU(N{-}1)$ $N_f{=}2N{-}2$
singularities and associated monodromies are reproduced by the $SU(N)$
$N_f{=}2N$ curve.  This allows us to check the gauge monodromies by
induction in $N$.  We need only compute for $SU(N)$ that monodromy not
contained in the Weyl group of $SU(N{-}1)$.  This is the ``special
monodromy'' identified in \cite{AF}, given in the scale-invariant
theory by
		\be
	{\bf S} = \pmatrix{\t P^{-1} & 0\cr 0 & P\cr},
		\label{weylmonod}
		\ee
where $P$ is the $(N{-}1)\times(N{-}1)$ matrix representation of
a $(1\ldots N)$ cyclic permutation.

For weak coupling, $|q|\ll 1$, and vevs much larger than the bare
masses $\phi_a \gg m_i$, the curve is approximately $y^2 =
\prod(x-\phi_a)^2 - q x^{2N}$.  Degenerations where two branch points
collide occur whenever $\phi_a=\phi_b$ up to corrections of order
$\sqrt q$, corresponding to the semiclassical positions of the gauge
singularities.  The special monodromies can be conveniently
measured by traversing a large circle in the $s_{N}$ complex plane,
fixing the other $s_\ell=0$, where the curve factorizes as
\be y^2 =
[(1-\sqrt q)x^{N}+ s_{N}]\cdot [(1+\sqrt q)x^{N}+ s_{N}].\ee
The branch
points are arranged in $N$ pairs with a pair at each $N$th root of
unity times $s_{N}^{1/N}$.  As $s_{N} \rightarrow e^{2\pi i} s_{N}$,
these pairs rotate into one another in a counter-clockwise sense.

Choose cuts and a basis for the cycles on the $SU(N)$ surface as shown
(for $SU(3)$) in Figure 1.  The intersection numbers for these cycles
are $\vev{\beta_a,\beta_b}=\vev{\gamma^a,\gamma^b}=0$ and
$\vev{\beta_a,\gamma^b}=-\delta_a^b + \delta_a^{b-1}$.  A canonical
basis of cycles is then $\beta_\ha$ and $\alpha^\ha \equiv
\sum_{a=1}^\ha \gamma^a$.  Note that $\beta_{N}$ is not independent of
the $\beta_\ha$'s: a simple contour deformation shows that $\sum_a
\beta_a=0$.  Similarly, $\sum_a \gamma^a=0$.  These relations hold
precisely when the residues of $\lambda$ on the $x$-plane sum to zero;
see the discussion preceding (\ref{curvformiii}).
\iffigs
\begin{figure}[hbtp]
\begin{center}
\leavevmode
\epsfxsize=8cm\epsfbox{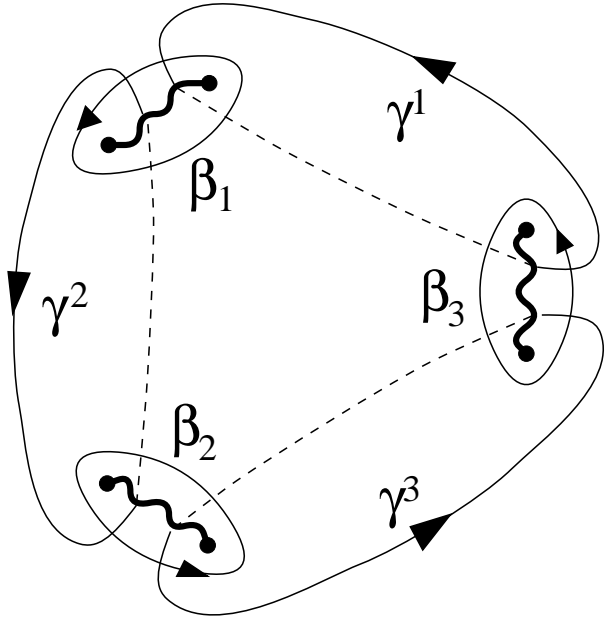}
\end{center}
\caption{Contours for a basis of cycles for the $SU(3)$ curve.
The thick wavy lines represent the cuts, solid contours are on the
first sheet, and dashed ones are on the second.}
\end{figure}
\fi

As $s_{N}\rightarrow e^{2\pi i}s_{N}$ the $\beta_a \rightarrow
\beta_{a+1}$, a cyclic permutation.  Using the fact that the $\beta_a$
sum to zero, this gives the $(N{-}1)\times(N{-}1)$ matrix representation
$P$ of the $(1\ldots N)$ permutation on the $\beta_\ha$ cycles, and
thus the $a^\ha$ periods.  Similarly, the monodromy along this path
takes $\gamma^a \rightarrow \gamma^{a+1}$ implying the $\alpha^\ha$
cycles (and thus the $a_D^\ha$ periods) transform by $\t P^{-1}$.
This is the classical monodromy predicted above.

This completes the induction step in the calculation of the gauge
singularity monodromies.  For the initial step we match our solution
and monodromies to the $SU(2)$ solution found in \cite{SWii}.  When
the bare masses are $m_i=(m,m,0,0)$ we find an explicit
$SL(2,\bC)$ transformation relating the two curves.  In addition,
the discriminants of the two curves factorize in the same way for
several other special cases, implying that the
positions and monodromies of the singularities in these cases
agree.  In fact, mimicking the argument of
\cite{SWii}, one can reverse this procedure to show that
these conditions fix the form of the $SU(2)$ curve to be
(\ref{curvformiii}).

The second class, referred to as quark singularities, occurs whenever
$\phi_a = -m_i$, and corresponds to the $q^i_a$, $\tq^a_i$
hypermultiplet becoming massless.  The massless quark can be taken to
have electric charge one with respect to a single $U(1)$ factor and to
carry quark numbers $n^j = \delta^j_1$.  The semiclassical monodromy
around the quark singularity can be read off from (\ref{dyonmon}).

Consider the curve at weak coupling
near a classical quark singularity, say $\phi_1+m_1\sim0$.
At weak coupling $g\mu+\mu_j \simeq m_j$
and for $x\sim \phi_1$ the
curve is approximately
\be y^2 = (x-\phi_1)^2 - 4qC (x+ m_1) \ee
where $C= \prod_{i>1} (x+m_i)/
\prod_{a>1} (x-\phi_a)$ is a slowly-varying function of $x$ and
$s_\ell$. This has a double zero at $x =
-m_1+qC$ for $\phi_1 = -m_1 - qC$, which is indeed near
the classical quark singularity for small $q$.
We can compute the
monodromy about this singularity by perturbing $\phi_1$ and computing
the locations of the two branch points from the above curve.  One
finds that as $\phi_1$ winds around the singular point the two branch
points are interchanged.  The
monodromy which follows from this is
nontrivial only in a $2\times 2$ block of (\ref{dyonmon}),
for which we find $\bS = {1\ 1\choose0\ 1}$, in agreement with the
semi-classical prediction.
But we find $\bT=0$, in seeming contradiction to the nonzero quark
number of the massless state.
However, recall that the periods are
defined only up to the residues of the form, since deforming the
contours to enclose one of the poles will shift the period, corresponding
to a physically unobservable redefinition of the global charges.  In the
case at hand it is natural to define the period $a_1$ by a
contour enclosing the pole at $-m_1$ which then gives
$\bT = \t(1 ~~ 0)$.

As this work was being completed we received \cite{HO} which addresses
related problems.

It is a pleasure to thank R. Donagi, M. Douglas, D. Friedan, K.
Intriligator, R.  Leigh, N. Seiberg, M. Strassler and E. Witten for
useful conversations.  The work of P.C.A. is supported by NSF grant
PHY92-45317 and by the Ambrose Monell Foundation.  M.R.P. is supported
in part by NSF grant PHY92-45317 and by the W.M. Keck Foundation.
A.D.S. is supported in part by DOE EPSCoR grant DE-FC02-91ER75661 and
by an Alfred P. Sloan Fellowship.

\end{document}

\end